\documentclass[final]{raa}

\usepackage{graphicx,times}             
\usepackage{natbib}
\usepackage{amssymb,amsmath}
\bibpunct{(}{)}{;}{a}{}{,}

\usepackage{ulem}

\newcommand\cts{counts~s$^{-1}$}

\usepackage[pagebackref=true]{hyperref}
\hypersetup{colorlinks = true, linkcolor = red, anchorcolor = red, citecolor = blue, filecolor = red, pagecolor = red, urlcolor = red}

\begin{document}

\title{Discrimination of background events in the PolarLight X-ray polarimeter}

\volnopage{Vol.0 (2021) No.0, 000--000}      
\setcounter{page}{1}          

\author{
Jiahuan Zhu\inst{1}, 
Hong Li\inst{1}, 
Hua Feng\inst{1,2}, 
Jiahui Huang\inst{2}, 
Xiangyun Long\inst{2}, 
Qiong Wu\inst{2}, 
Weichun Jiang\inst{3}, 
Massimo Minuti\inst{4}, 
Saverio Citraro\inst{4}, 
Hikmat Nasimi\inst{4}, 
Dongxin Yang\inst{2}, 
Jiandong Yu\inst{5}, 
Ge Jin\inst{6}, 
Ming Zeng\inst{2}, 
Peng An\inst{5}, 
Luca Baldini\inst{4}, 
Ronaldo Bellazzini\inst{4}, 
Alessandro Brez\inst{4}, 
Luca Latronico\inst{7}, 
Carmelo Sgr\`{o}\inst{4}, 
Gloria Spandre\inst{4}, 
Michele Pinchera\inst{4}, 
Fabio Muleri\inst{8}, 
Paolo Soffitta\inst{8}, 
Enrico Costa\inst{8}
}

\institute{
Department of Astronomy, Tsinghua University, Beijing 100084, China; {\it lihong2019@tsinghua.edu.cn,hfeng@tsinghua.edu.cn}\\
\and
Department of Engineering Physics, Tsinghua University, Beijing 100084, China\\
\and
Key Laboratory for Particle Astrophysics, Institute of High Energy Physics, Chinese Academy of Sciences, Beijing 100049, China\\
\and
INFN-Pisa, Largo B. Pontecorvo 3, 56127 Pisa, Italy\\
\and
School of Electronic and Information Engineering,  Ningbo University of Technology, Ningbo, Zhejiang 315211, China\\
\and
North Night Vision Technology Co., Ltd., Nanjing 211106, China\\
\and
INFN, Sezione di Torino, Via Pietro Giuria 1, I-10125 Torino, Italy\\
\and
IAPS/INAF, Via Fosso del Cavaliere 100, 00133 Rome, Italy\\
\vs\no
{\small Received~~2021 month day; accepted~~2021~~month day}}

\abstract{
PolarLight is a space-borne X-ray polarimeter that measures the X-ray polarization via electron tracking in an ionization chamber.  It is a collimated instrument and thus suffers from the background on the whole detector plane. The majority of background events are induced by high energy charged particles and show ionization morphologies distinct from those produced by X-rays of interest. Comparing on-source and off-source observations, we find that the two datasets display different distributions on image properties. The boundaries between the source and background distributions are obtained and can be used for background discrimination. Such a means can remove over 70\% of the background events measured with PolarLight. This approaches the theoretical upper limit of the background fraction that is removable and justifies its effectiveness. For observations with the Crab nebula, the background contamination decreases from 25\% to 8\% after discrimination, indicative of a polarimetric sensitivity of around 0.2~Crab for PolarLight. This work also provides insights into future X-ray polarimetric telescopes.
}

\authorrunning{J. Zhu, H. Li, H. Feng, et al}            
\titlerunning{Discrimination of background events in PolarLight}  
   
\maketitle

\section{Introduction}
\label{sec:intro}

PolarLight is a miniature space program dedicated to astrophysical X-ray polarimetry in the energy range of 2--8 keV~\citep{Feng2019, Feng2020}. It features the gas pixel detector (GPD), which is a gas-filled ionization chamber allowing us to track the trajectory of photoelectrons ejected by incident X-rays~\citep{Costa2001}.  The polarization of the X-ray source can thus be measured via the distribution of photoelectron emission angles~\citep{Bellazzini2013}.  

Since the launch in October, 2018, PolarLight has been serving in the space for more than two years~\citep{Li2021}.  The science targets of PolarLight include the Crab nebula~\citep{Feng2020a}, Sco X-1, and some transient X-ray binaries that are comparably bright. In the meanwhile, background observations are conducted when the source is occulted by the Earth.  A particle tracking simulation of the in-orbit radiation environment with a mass model of the whole satellite suggests that the measured background in the energy band of 2--8 keV is mainly induced by charged particles, with a contribution of $\sim$76\% from high energy electrons/positrons, $\sim$17\% from high energy protons, and $\sim$7\% from cosmic X-rays leaking through surrounding materials~\citep{Huang2021}.   

The background level is an important factor determining the sensitivity of polarization measurement~\citep{Weisskopf2010}.  Background removal is particularly important for PolarLight because it is a collimated instrument, where the background all over the detector plane needs be taken into account.  As an ionization chamber, PolarLight has the capability of distinguishing particle types and removing part of the background events based on the track images. \citet{Huang2021} reveal that the majority of background events yield long, straight tracks that are distinct from those produced by X-rays in the energy range of our interest. However, they could also produce secondary electrons in the PolarLight energy band, leading to energy deposits and track images indistinguishable from those due to source photons.  It is found that roughly 28\% of the measured background appear like X-ray events, while 72\% of them are typical of high energy charged particles.  This implies that an effective discrimination algorithm can remove up to $\sim$72\% of the background events in the energy range of 2--8 keV.  In this paper, we propose and justify that a simple and robust means can effectively fulfill the purpose. 

\section{Discrimination method}
\label{sec:method}

When there are energy deposits in the detector, multiple pixels could be triggered as long as the charge collected on the pixel exceeds a pre-defined, constant level in the electronics. Then, a rectangular readout window is determined by adding 8 and 10 pixels along $X$ and $Y$, respectively, on each side of the triggered pixels. The readout window may extend to the very end pixels, and consequently, events near the edge may have a smaller number of margin pixels. Therefore, an event refers to an image of charges in the readout window following an electronic trigger. 

\begin{figure}[b]
\centering
\includegraphics[width=0.4\columnwidth]{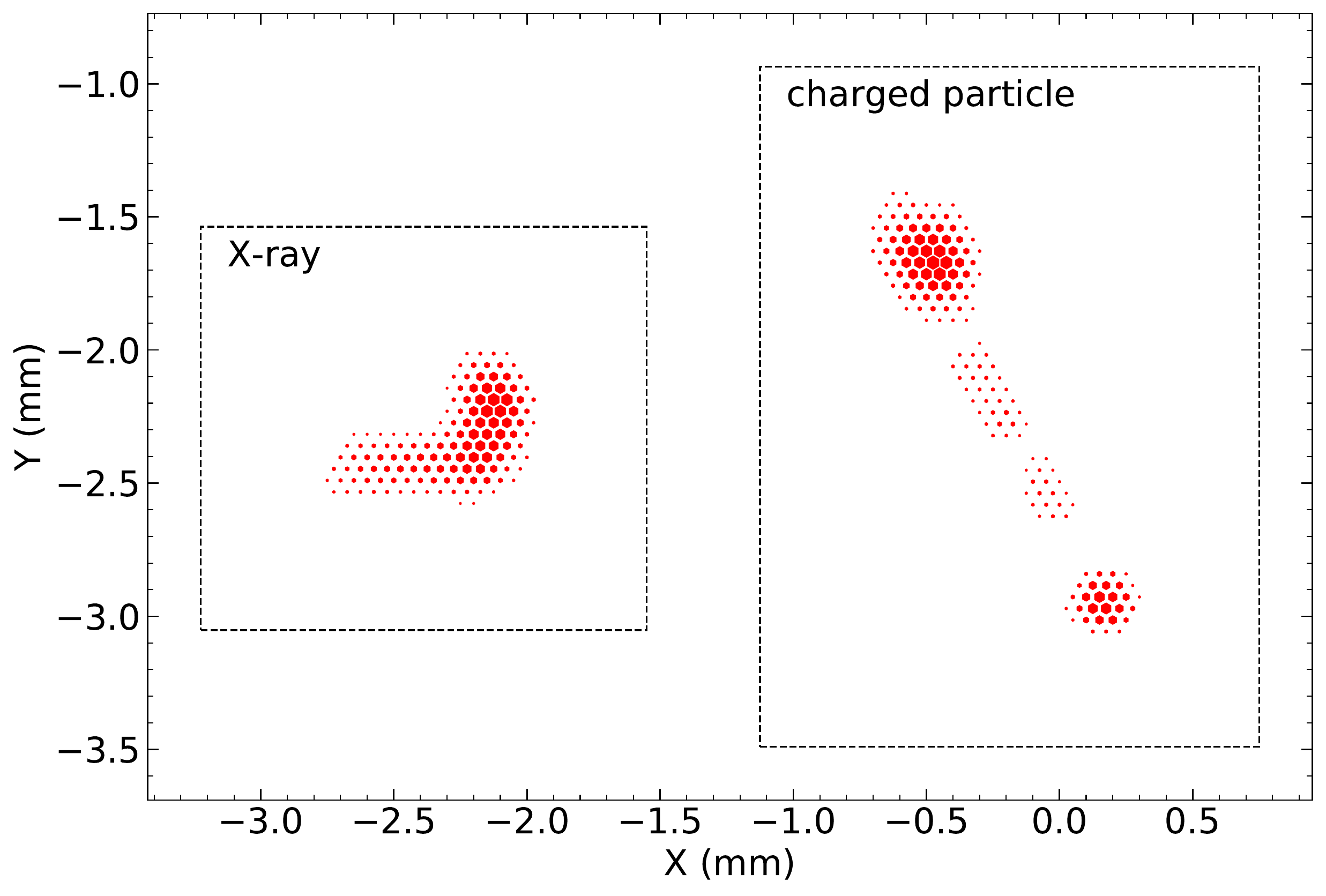}
\caption{Typical event images in the energy bin of 5.0--5.5~keV likely produced by a source photon and a background charged particle, respectively. The dashed box is the readout window for each event. }
\label{fig:track}
\end{figure}

The raw image consists of both signals and electronic noise. Before imaging analysis, we employ a noise cut and set the pixel values to zero if they are below the noise threshold. Thanks to charge magnification with a gain from a few hundred to a few thousand in the detector,  the electronic noise can be easily cut off.  However, the gain varies with time, and the choice of the noise threshold may affect the wing size of the signal image. To minimize such effects, we adjust the noise threshold according to the detector gain, so that the image after noise cut remains the same at various gains.  The threshold is calculated for each observation and is uniform across the detector plane.

Two event images that are likely resulted from an X-ray and a high energy charged particle, respectively, are displayed in Figure~\ref{fig:track} for comparison. The dashed box indicates the readout window.  The two events are selected to have similar energies.  The X-ray-like event shows a curved shape, while the comic-ray-like event is typically straight.  Also, cosmic-ray-like events tend to spread on a larger number of pixels, due to a relatively low rate of energy loss.  Such differences allow us to distinguish them. 

\begin{figure}[t]
\centering
\includegraphics[width=0.4\columnwidth]{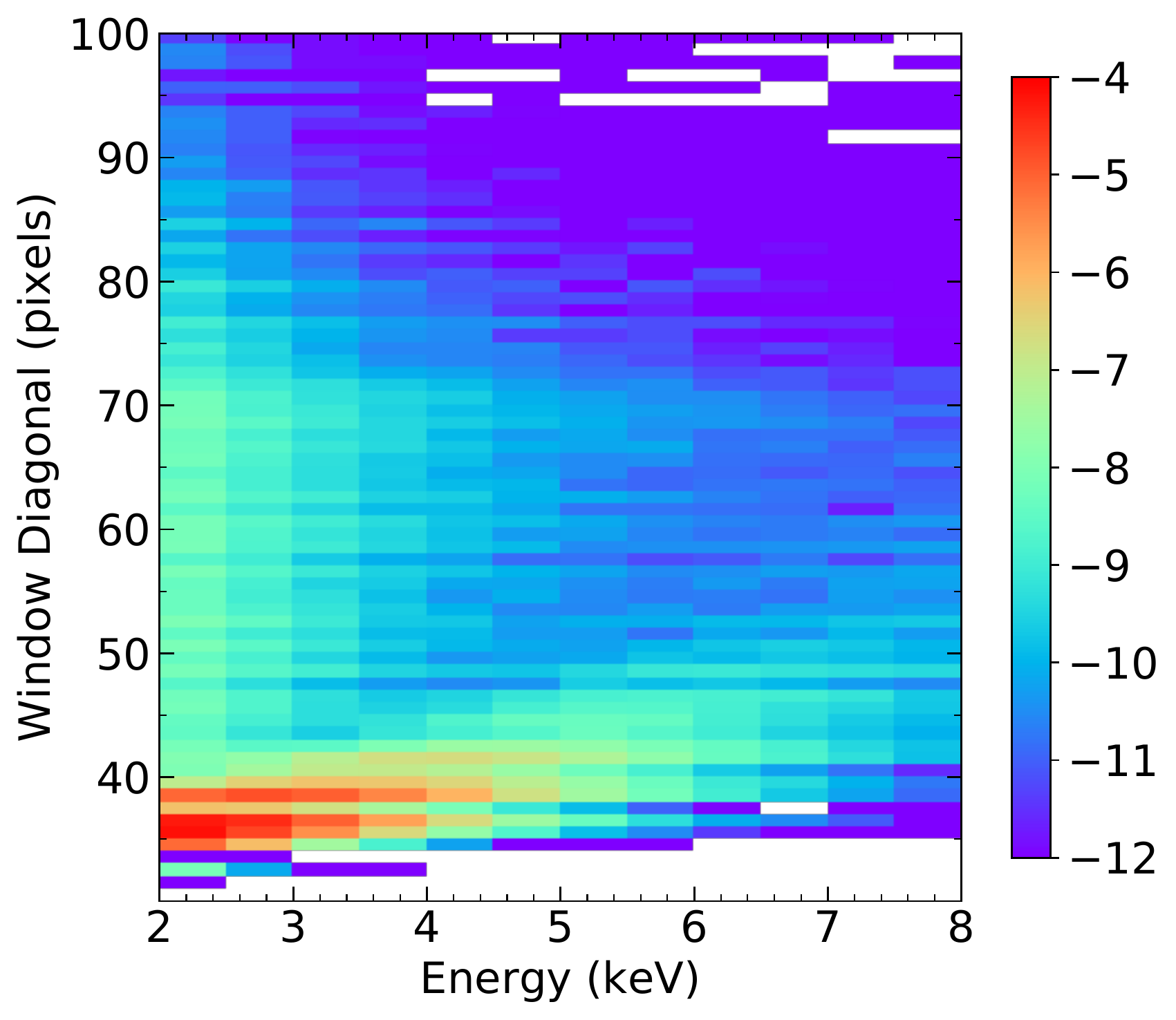}
\includegraphics[width=0.4\columnwidth]{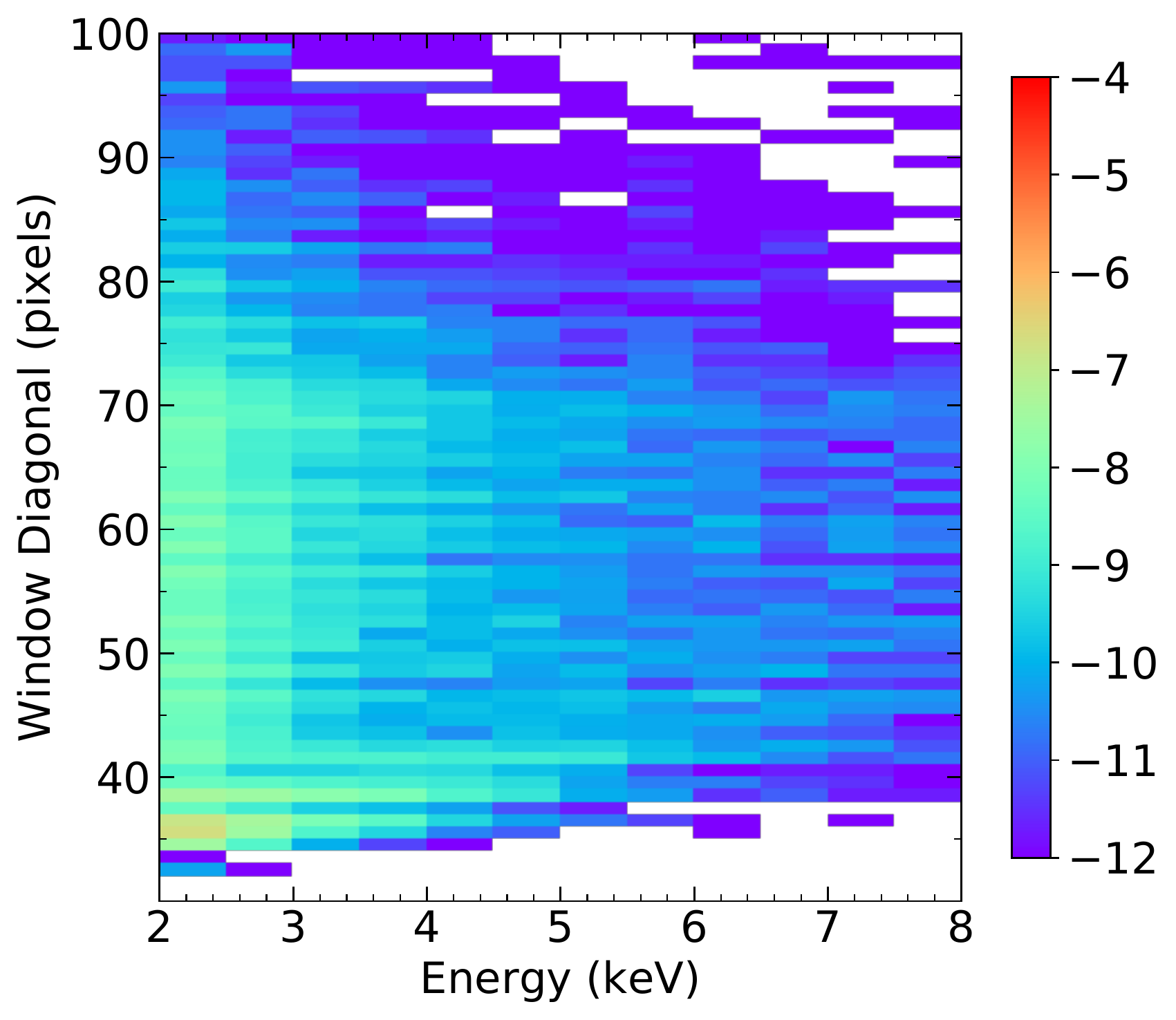}
\caption{Histogram of events as a function of energy ($E$) and the readout window diagonal size ($L$), for observations of the Crab nebula (\textbf{left}) and background (\textbf{right}). The color indicates the logarithmic count rate in units of \cts\ in each ($\Delta E$, $\Delta L$) pixel.}
\label{fig:le}
\end{figure}

\begin{figure}[t]
\centering
\includegraphics[width=0.4\columnwidth]{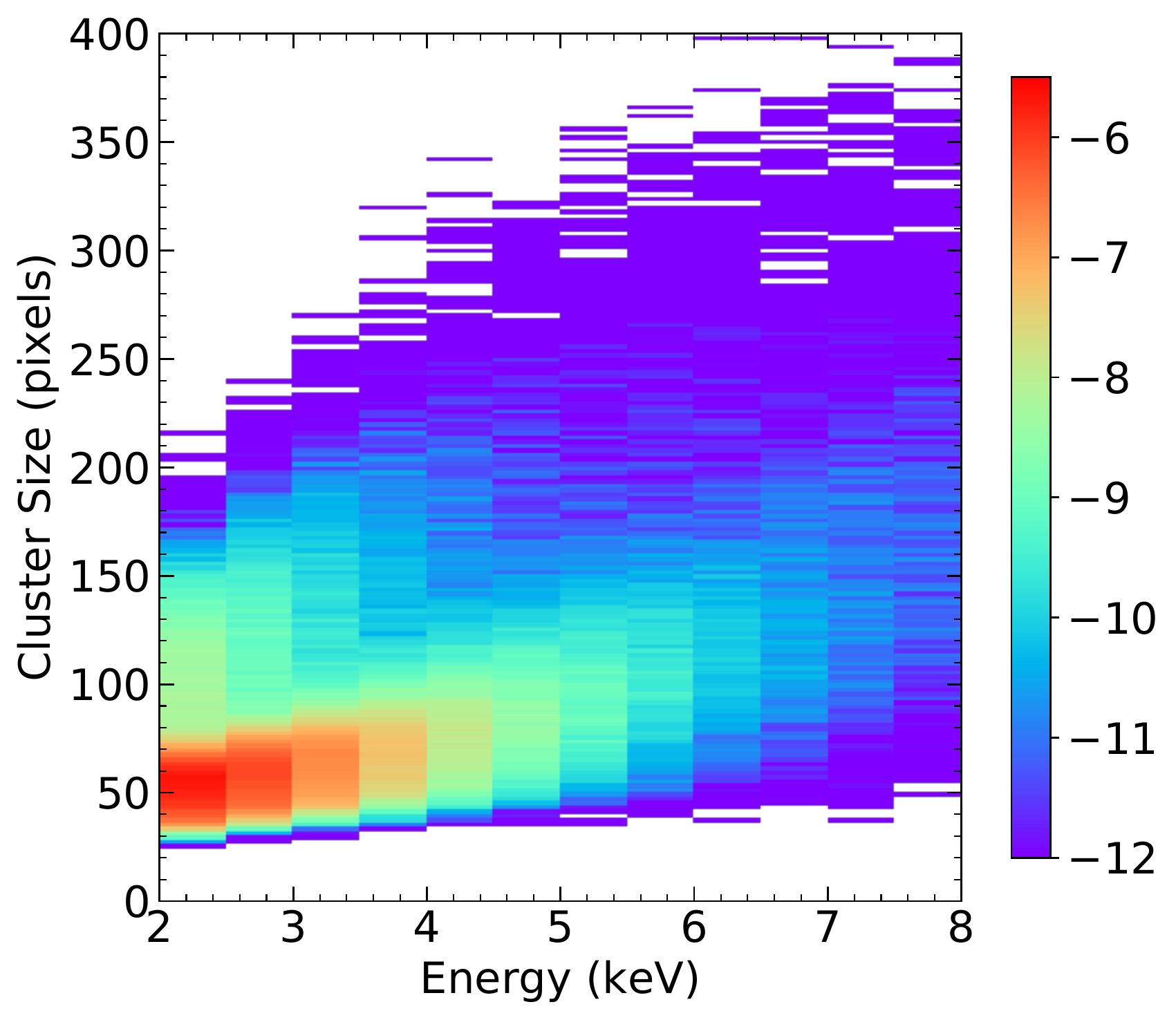}
\includegraphics[width=0.4\columnwidth]{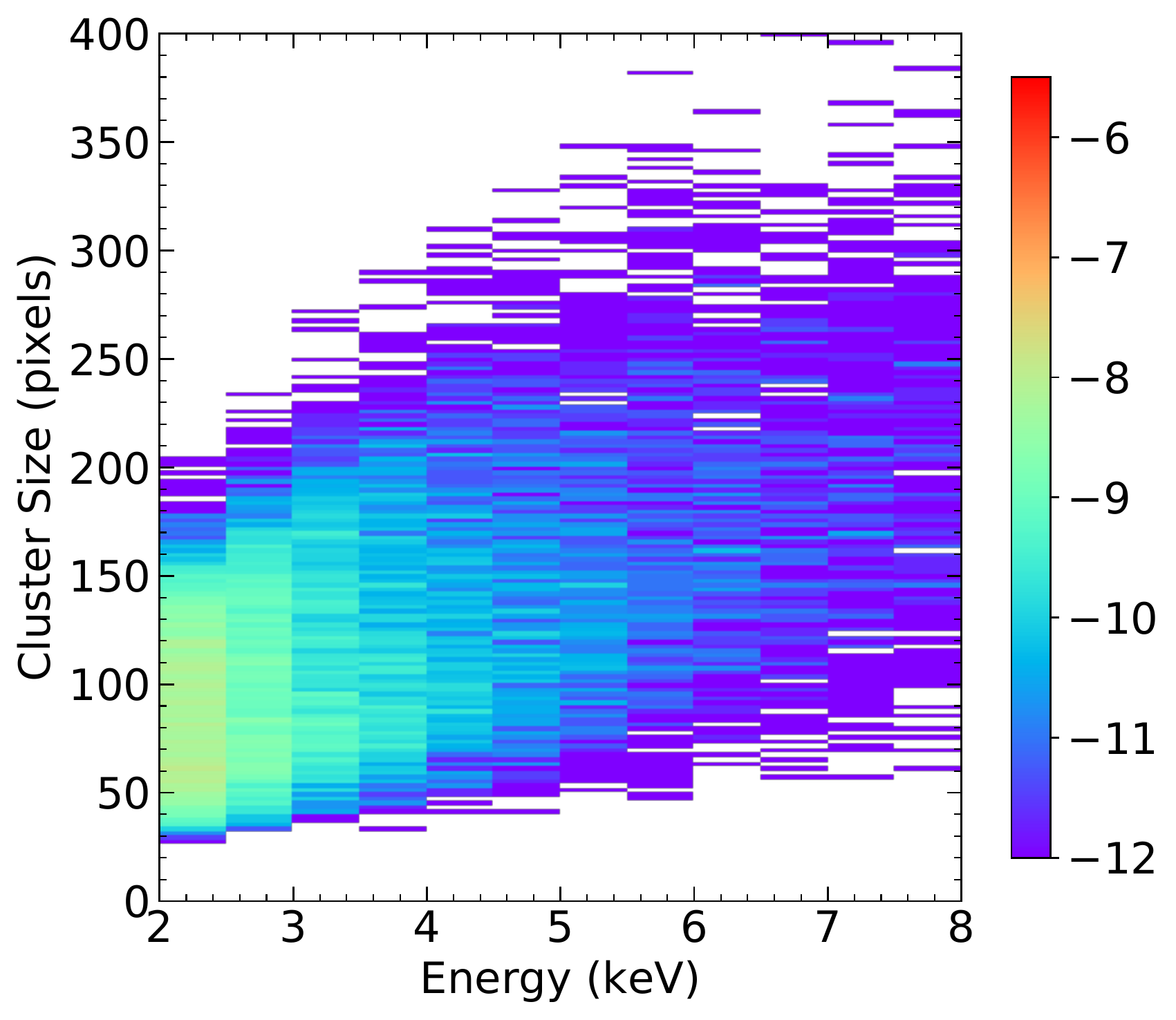}
\caption{Histogram of events as a function of energy ($E$) and the cluster size ($S$), for observations of the Crab nebula (\textbf{left}) and background (\textbf{right}). The color indicates the logarithmic count rate in units of \cts\ in each ($\Delta E$, $\Delta S$) pixel.}
\label{fig:se}
\end{figure}

In this work, we use all of the observations of the Crab nebula with a total exposure of about 1.4~Ms, and all the background observations with a total exposure of about 390~ks.  The energy calibration is performed following \citet{Li2021} by comparing the model and measured spectra. We select events in the energy range of 2--8 keV and in the central $\pm7$~mm region on the detector plane (to exclude those near the edge with an incomplete track image).  For the Crab data,  observations within an off-axis angle of 0.2$^\circ$ are selected. For each event, we extract its energy ($E$) and two characteristics from the image, the diagonal size ($L$) of the readout window and the cluster size ($S$) that represents the number of pixels above the noise threshold. The 2D distribution of events on the $L-E$ plane is plotted in Figure~\ref{fig:le} for observations of the Crab nebula, which include both source and background events, and observations of the pure background. Similarly, the distributions on the  $S-E$ plane are plotted in Figure~\ref{fig:se}.

\begin{figure}[t]
\centering
\includegraphics[width=0.4\columnwidth]{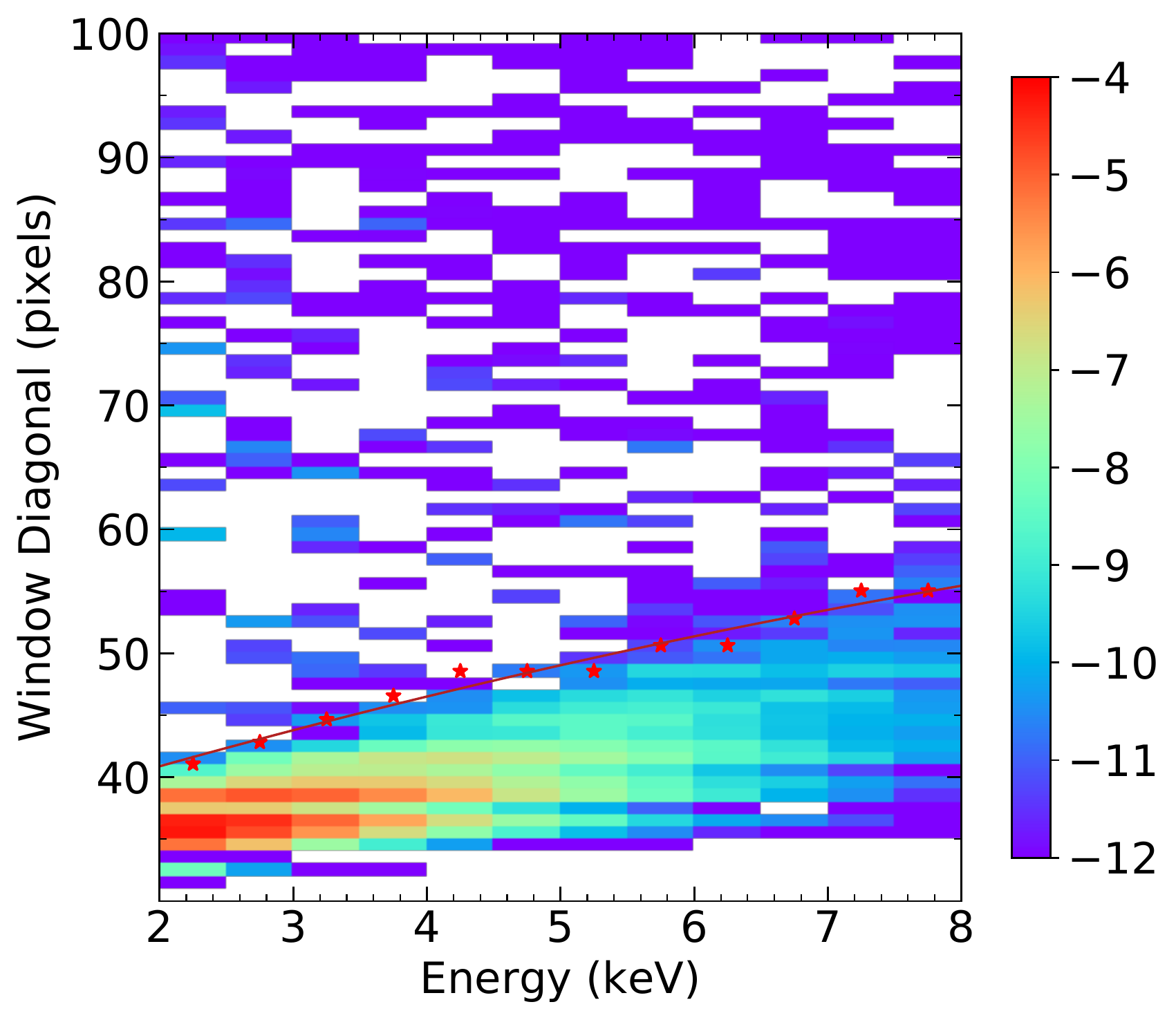}
\includegraphics[width=0.4\columnwidth]{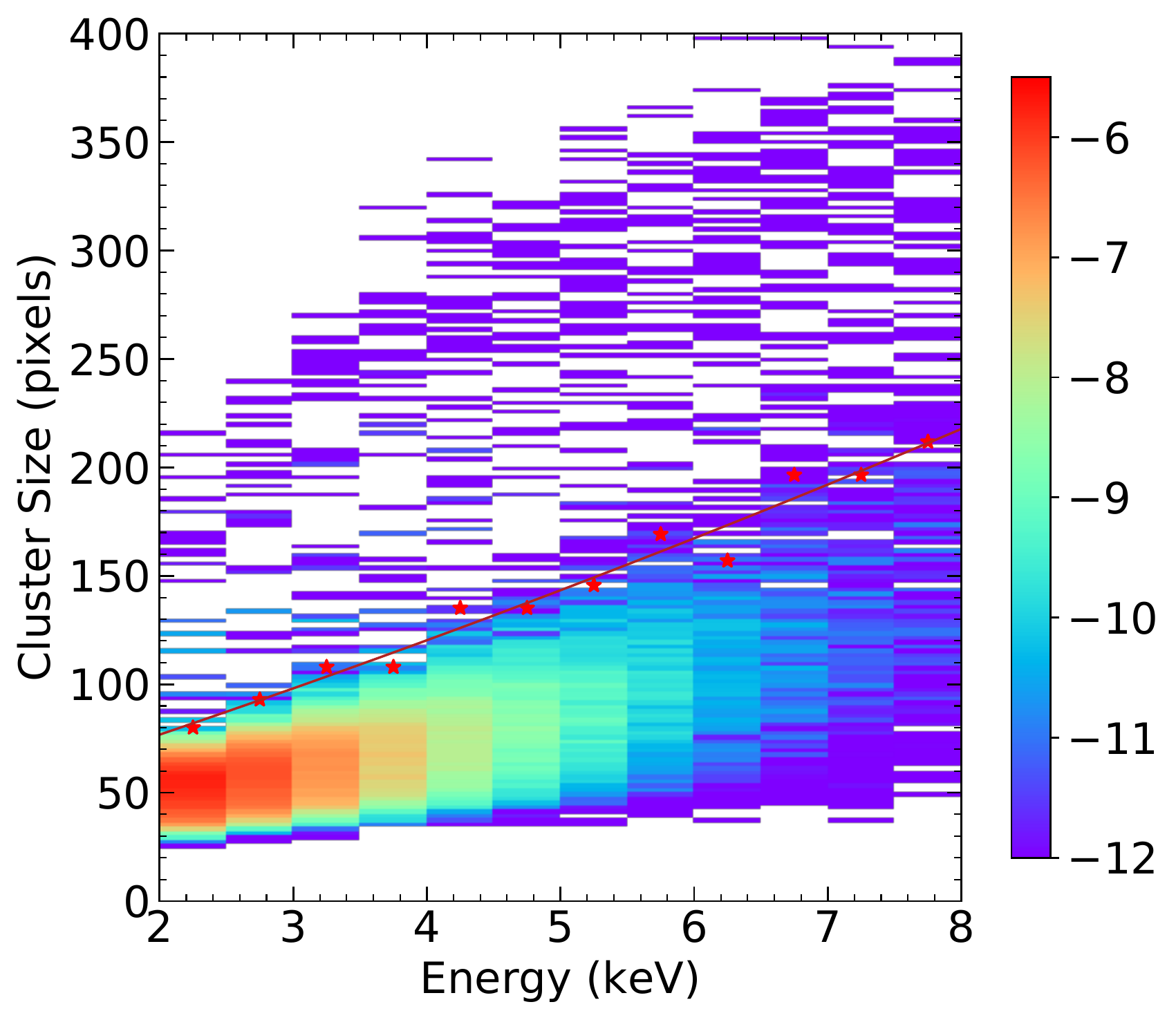}
\caption{Pure source distribution on the $L-E$ (\textbf{left}) and $S-E$ (\textbf{right}) planes. The color indicates the logarithmic count rate in units of \cts\ in each ($\Delta E$, $\Delta L$) or ($\Delta E$, $\Delta S$) pixel. The red stars are the calculated boundaries between the source-dominant and background-dominant regions in each energy bin, and the red curves are the discrimination curves. }
\label{fig:pure_src}
\end{figure}

For both distributions, it is obvious that the on-source distributions contain two components, one clustering toward the low $L$ and low $S$ end on top of an extended component similar to the off-source (background) distribution.  This is consistent with the fact that background events tend to have large $L$ and $S$ compared with source events. Then, we subtract the off-source distribution from the on-source distribution, such that the pure source distribution is seen (see Figure~\ref{fig:pure_src}).  

\begin{figure*}[h!]
\centering
\includegraphics[width=0.47\textwidth]{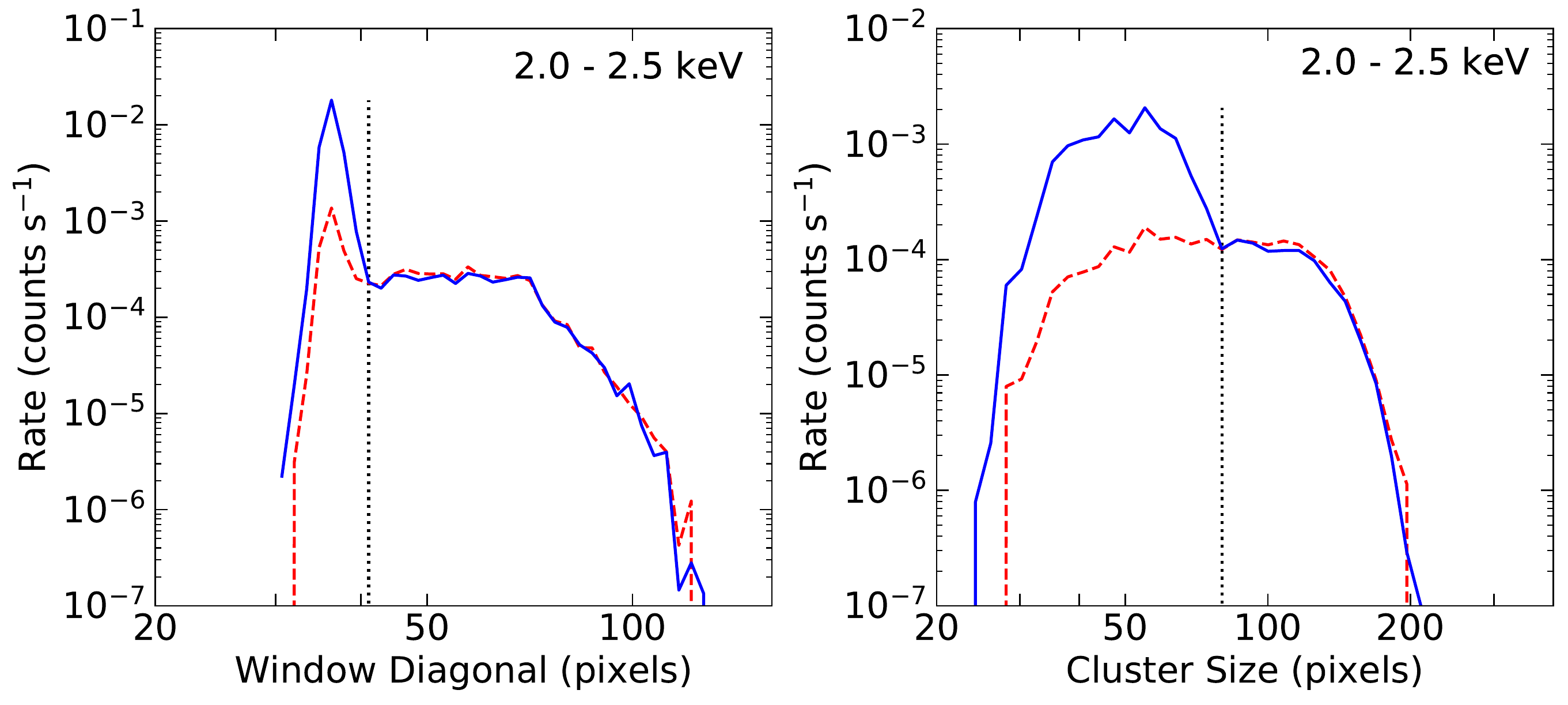}
\includegraphics[width=0.47\textwidth]{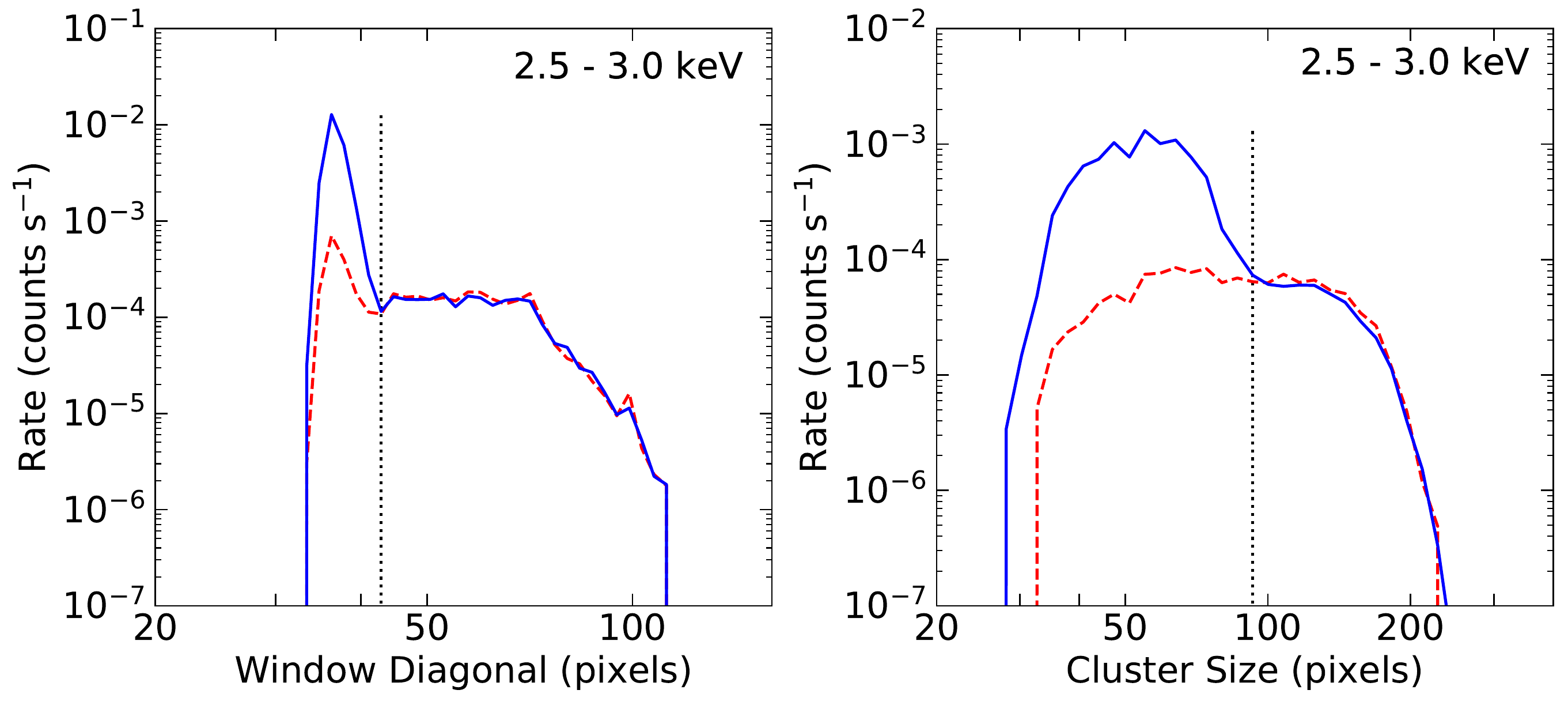}
\includegraphics[width=0.47\textwidth]{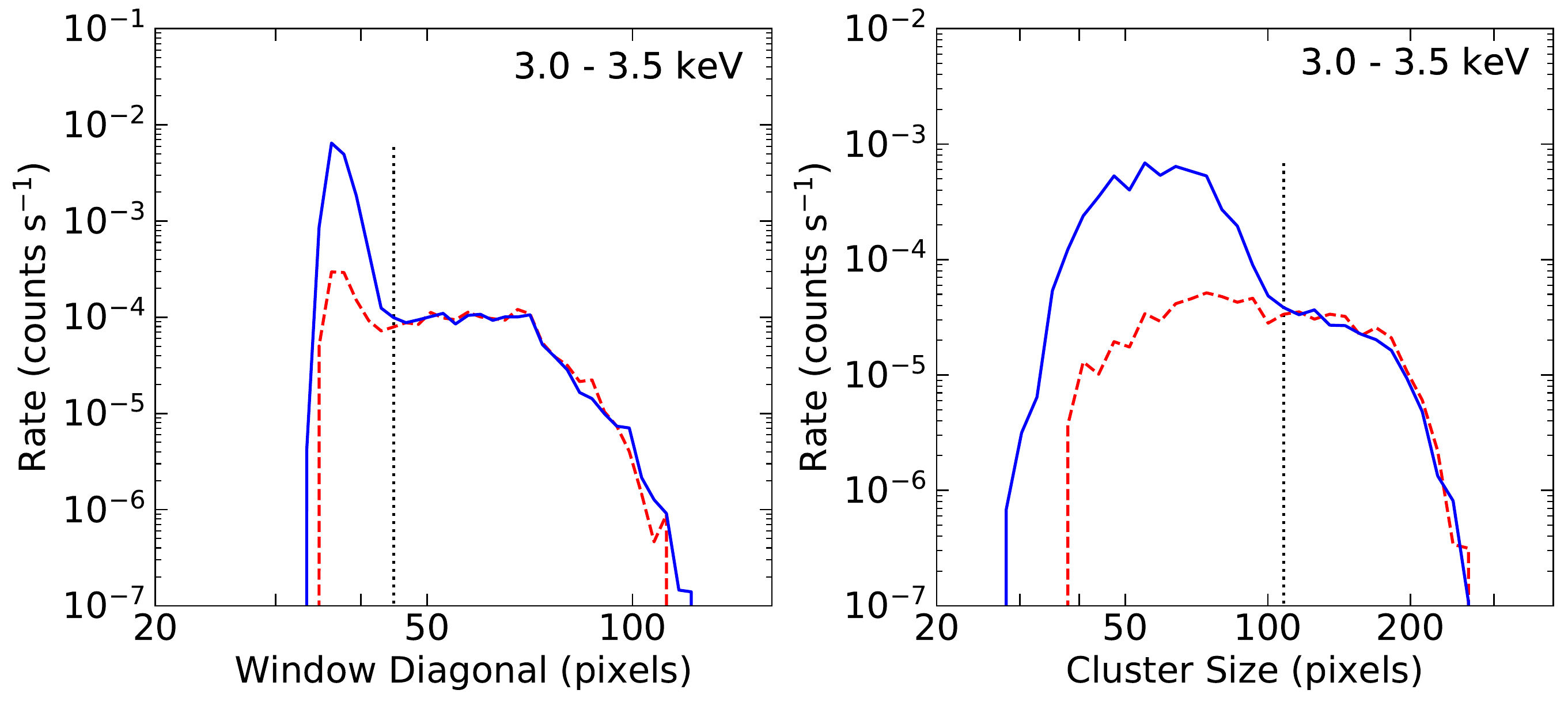}
\includegraphics[width=0.47\textwidth]{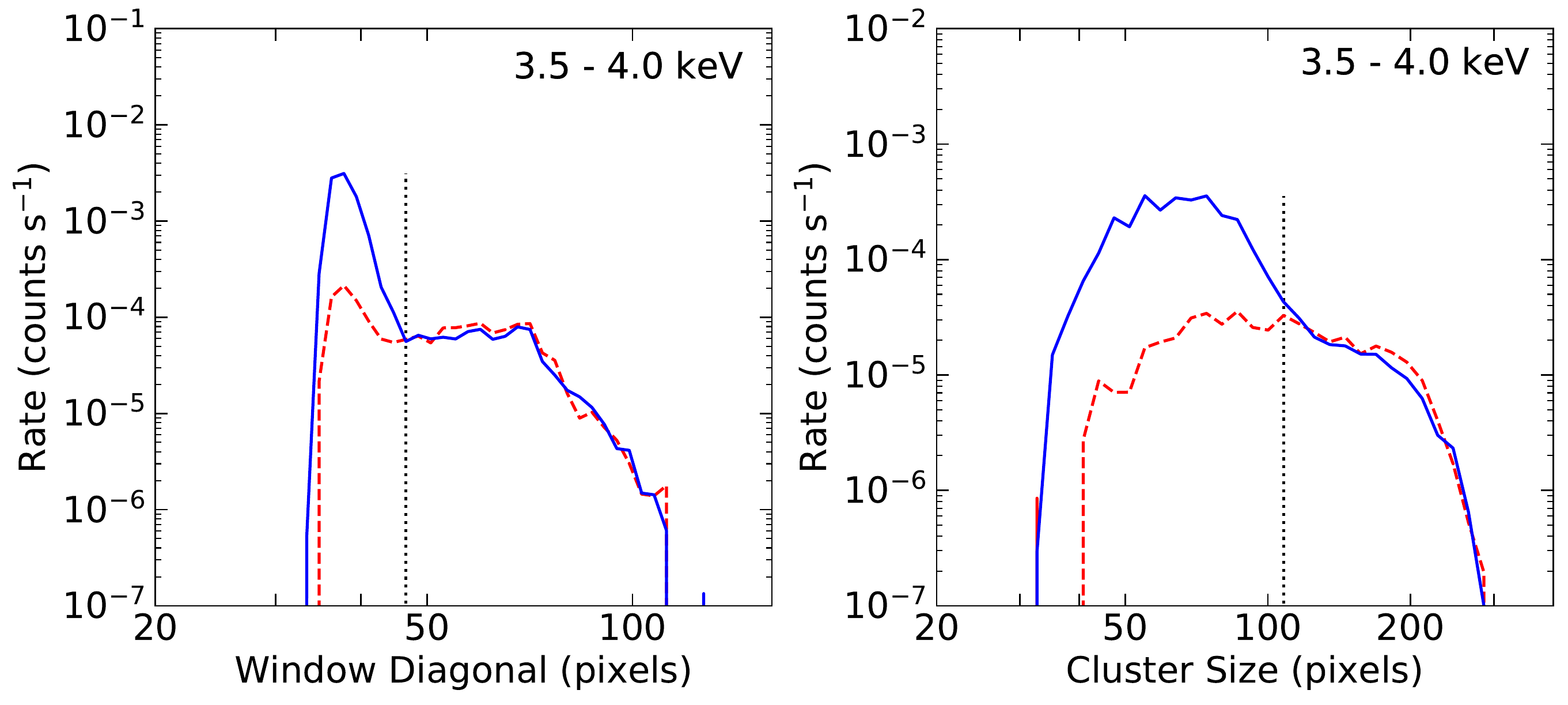}
\includegraphics[width=0.47\textwidth]{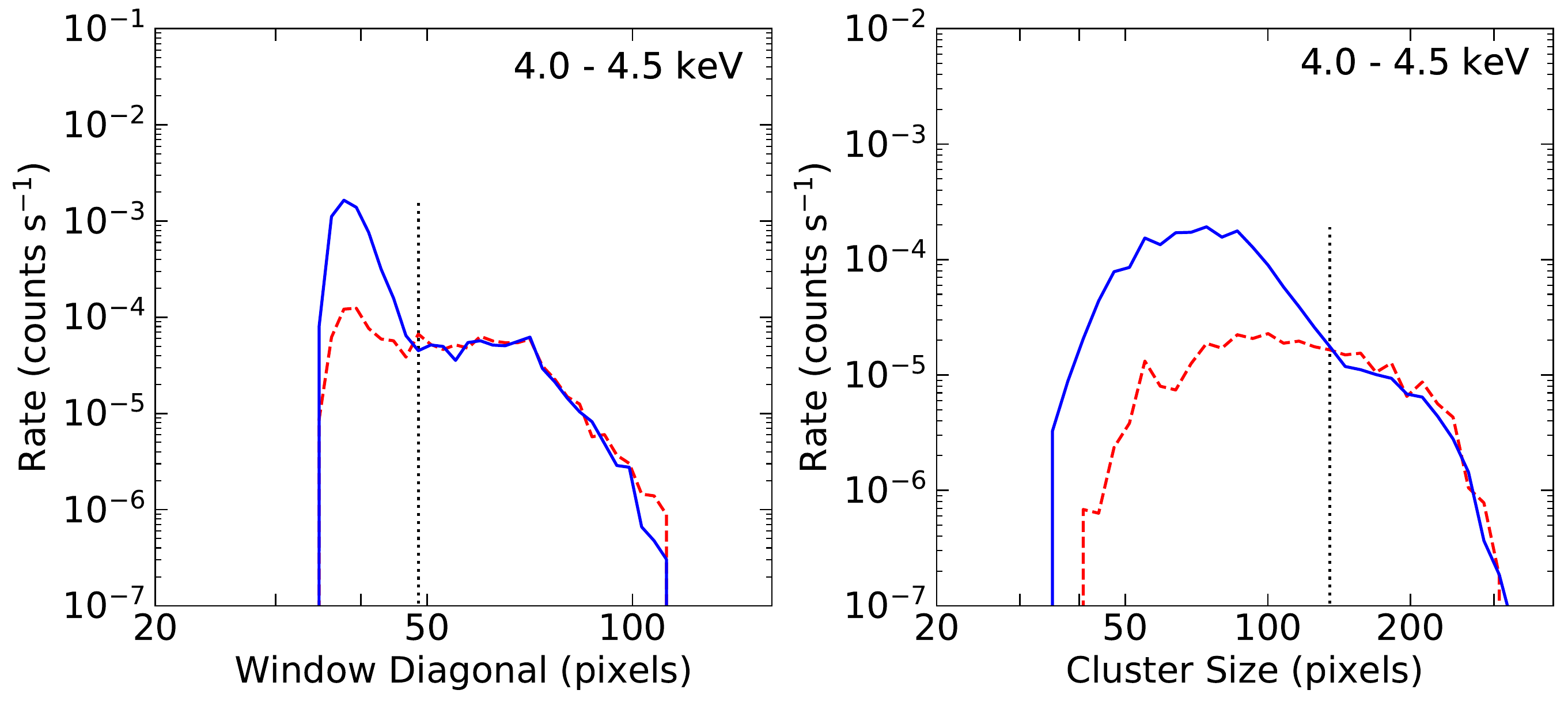}
\includegraphics[width=0.47\textwidth]{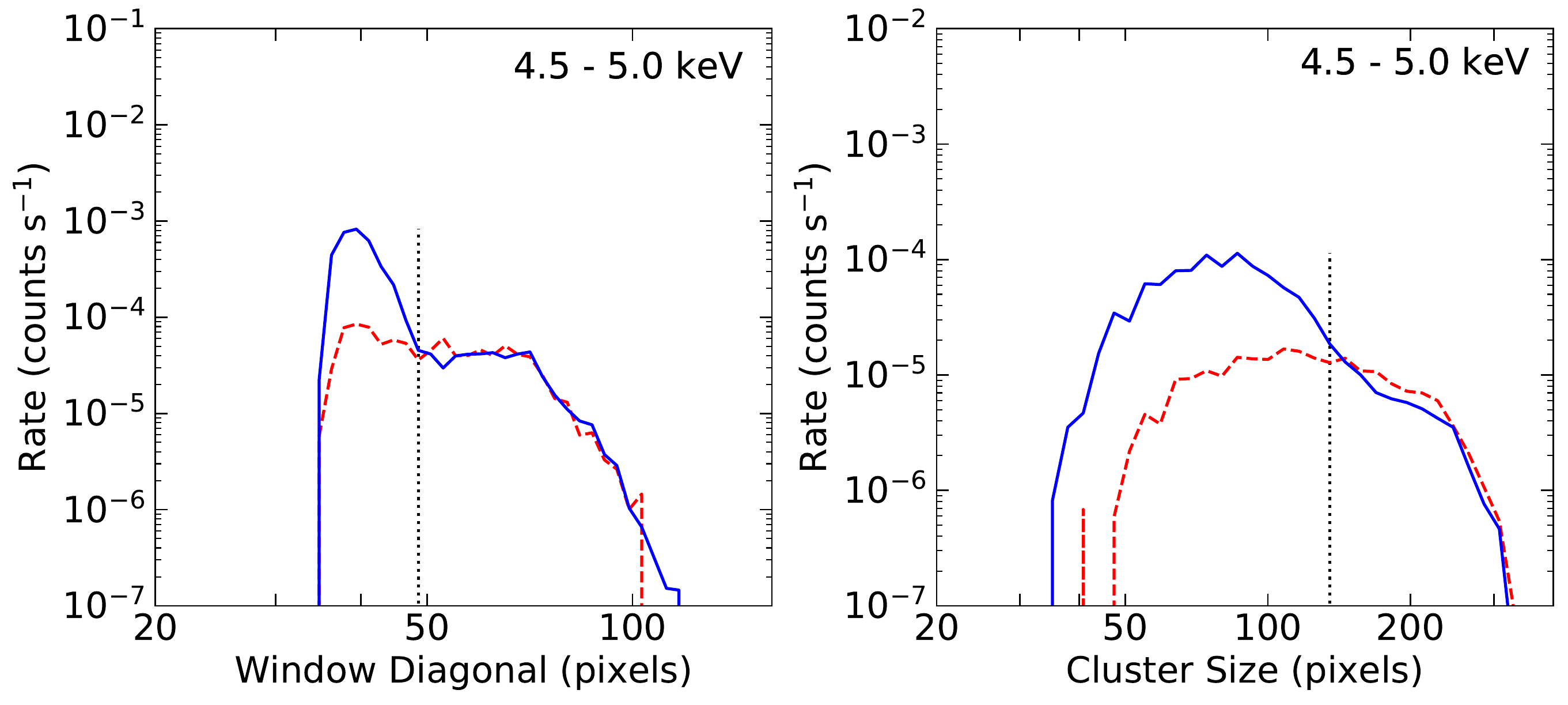}
\includegraphics[width=0.47\textwidth]{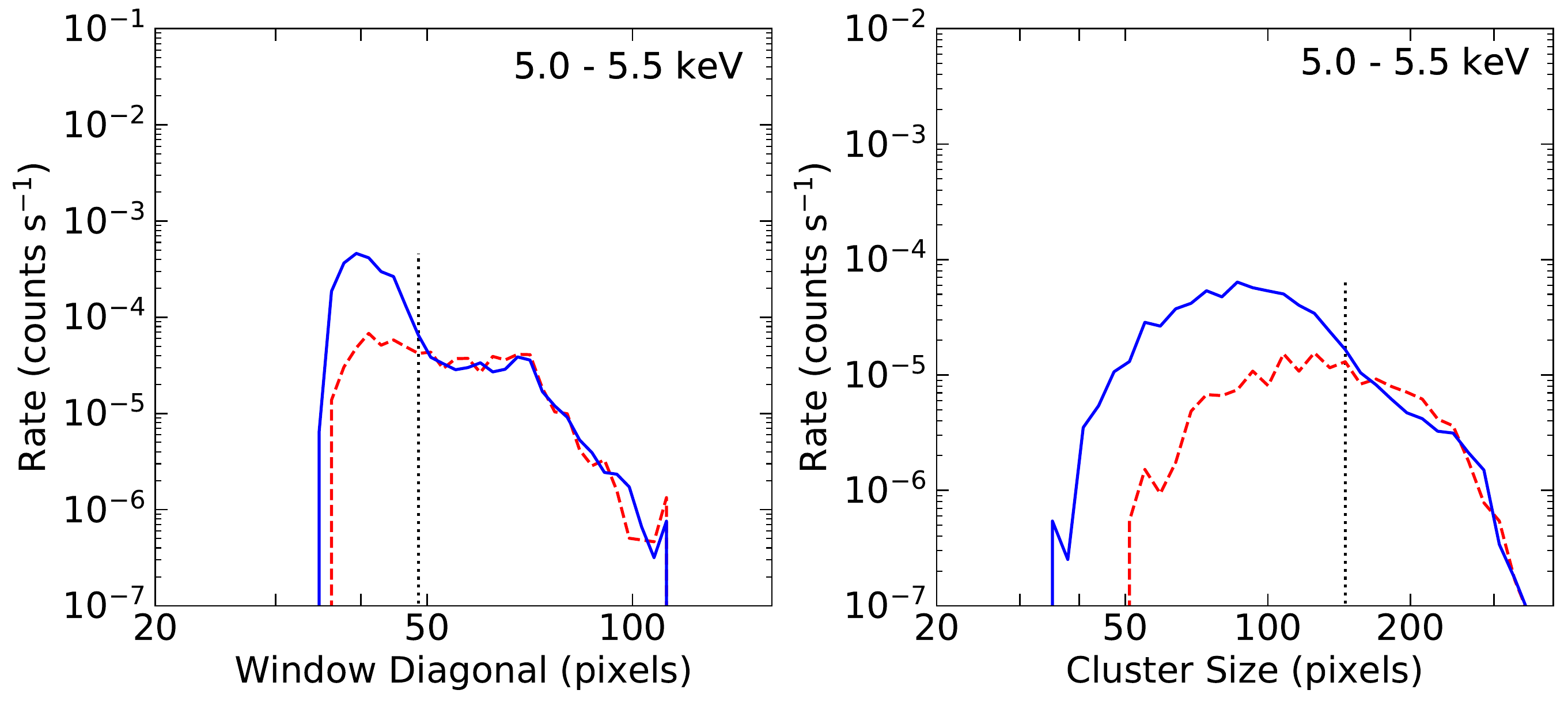}
\includegraphics[width=0.47\textwidth]{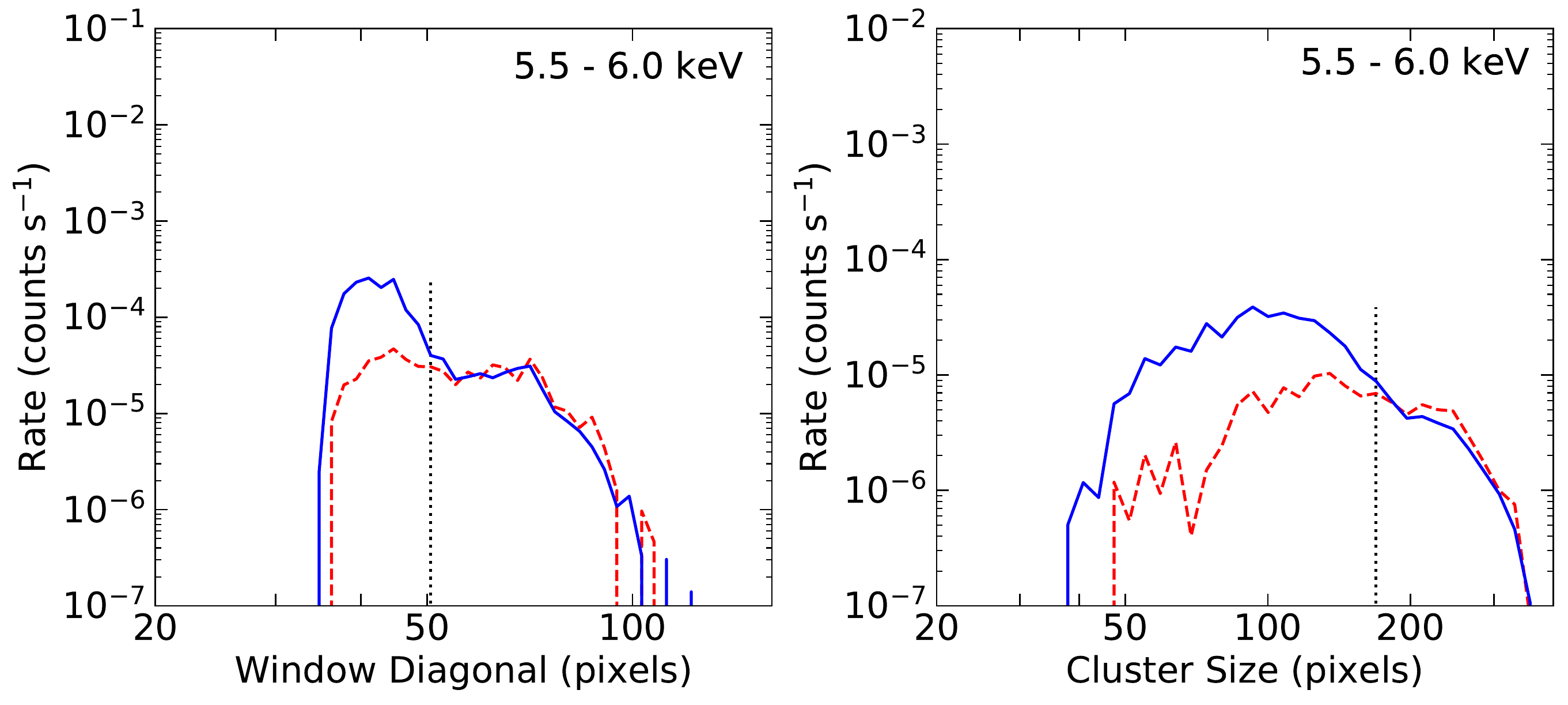}
\includegraphics[width=0.47\textwidth]{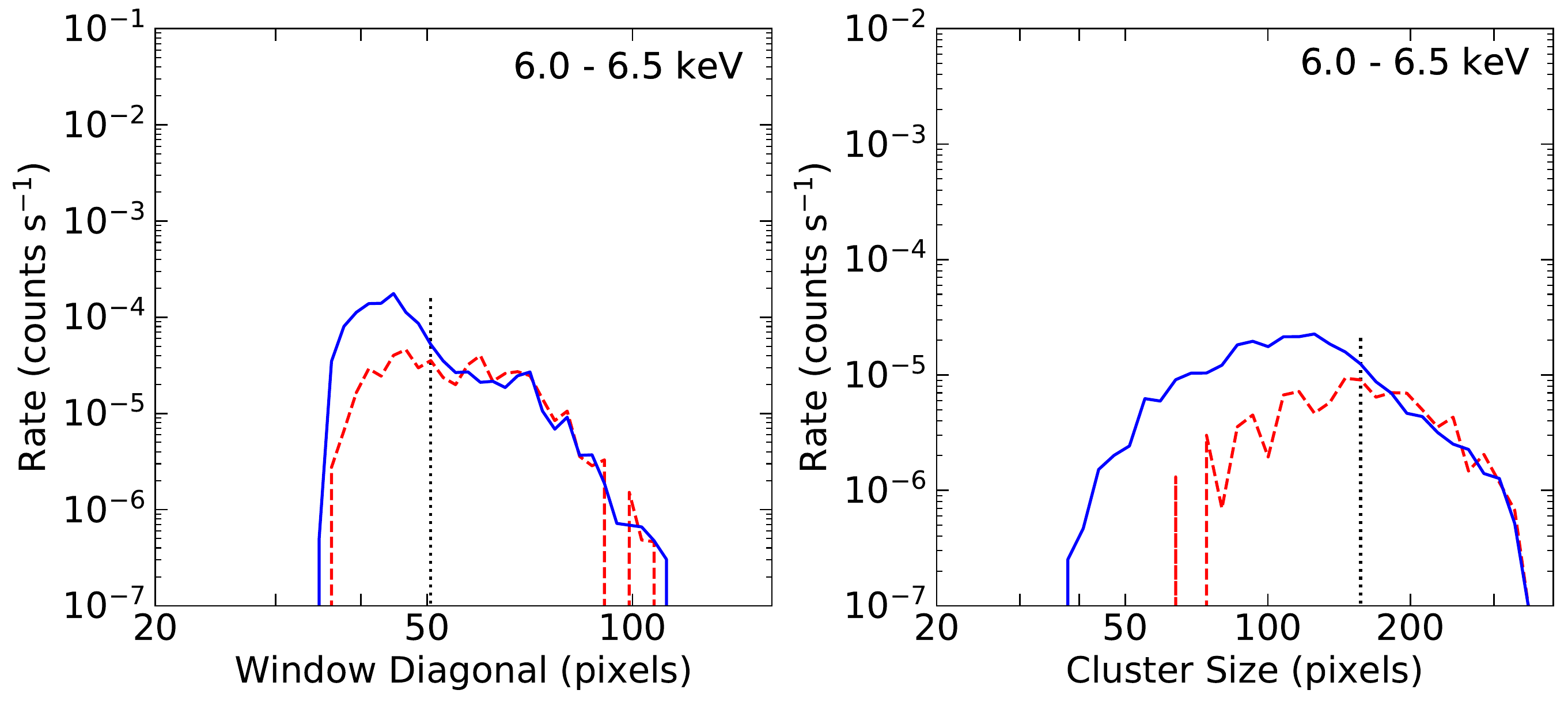}
\includegraphics[width=0.47\textwidth]{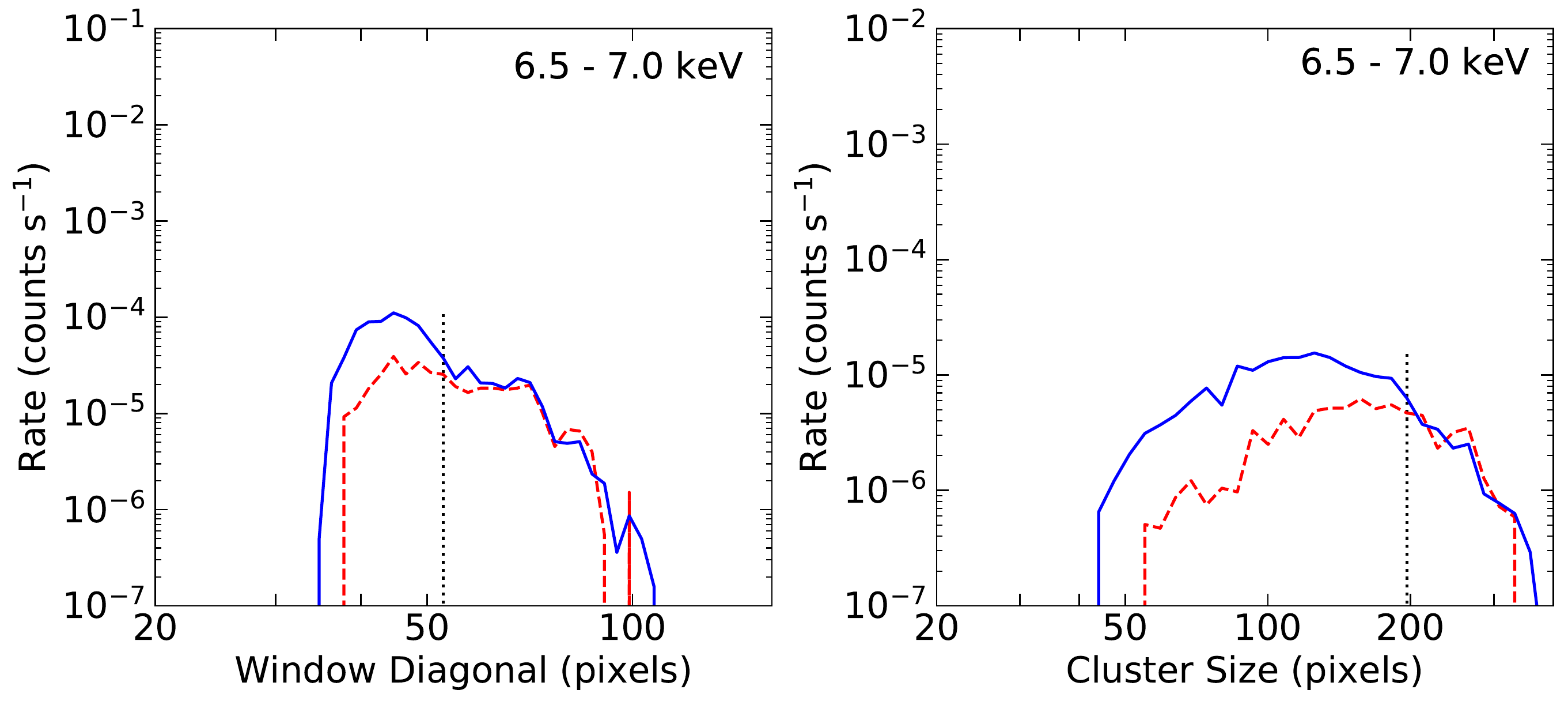}
\includegraphics[width=0.47\textwidth]{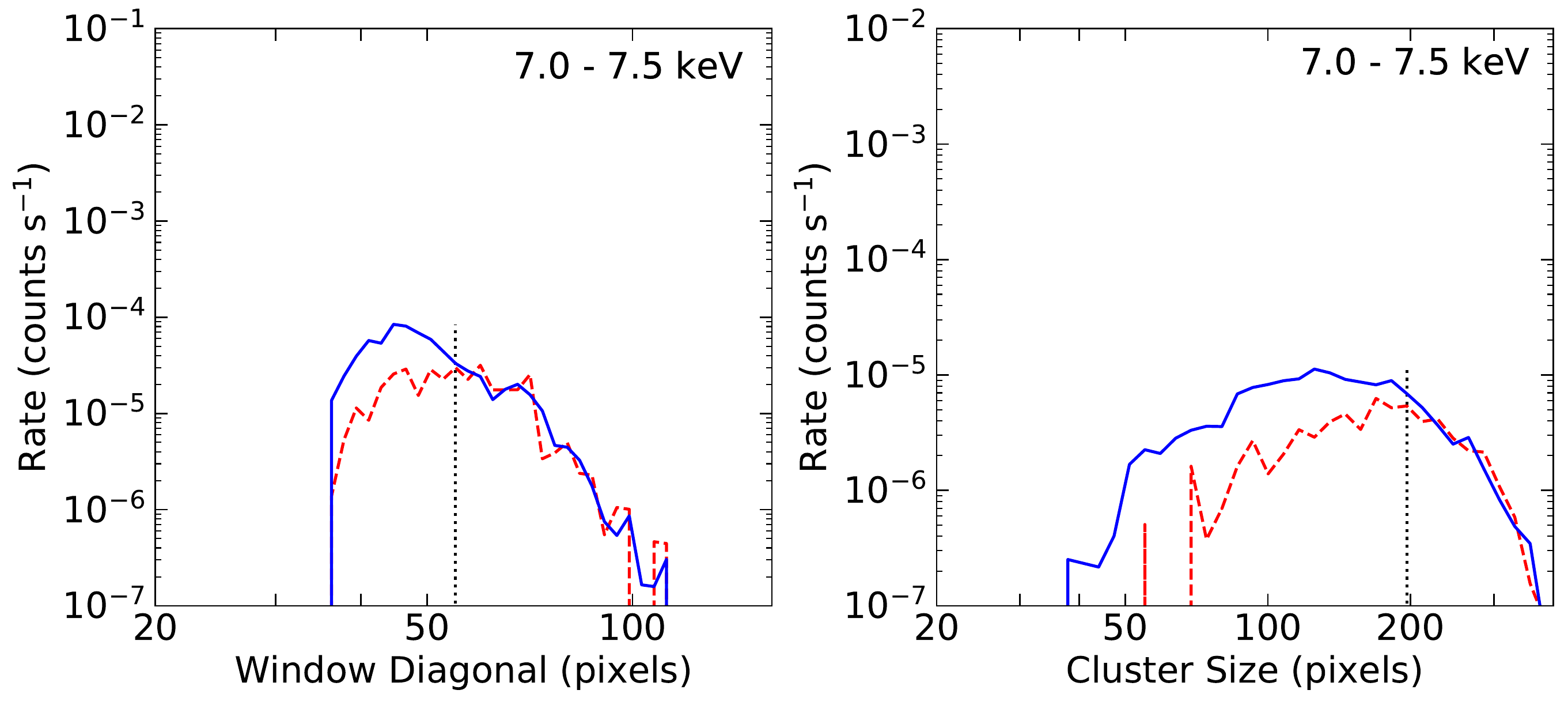}
\includegraphics[width=0.47\textwidth]{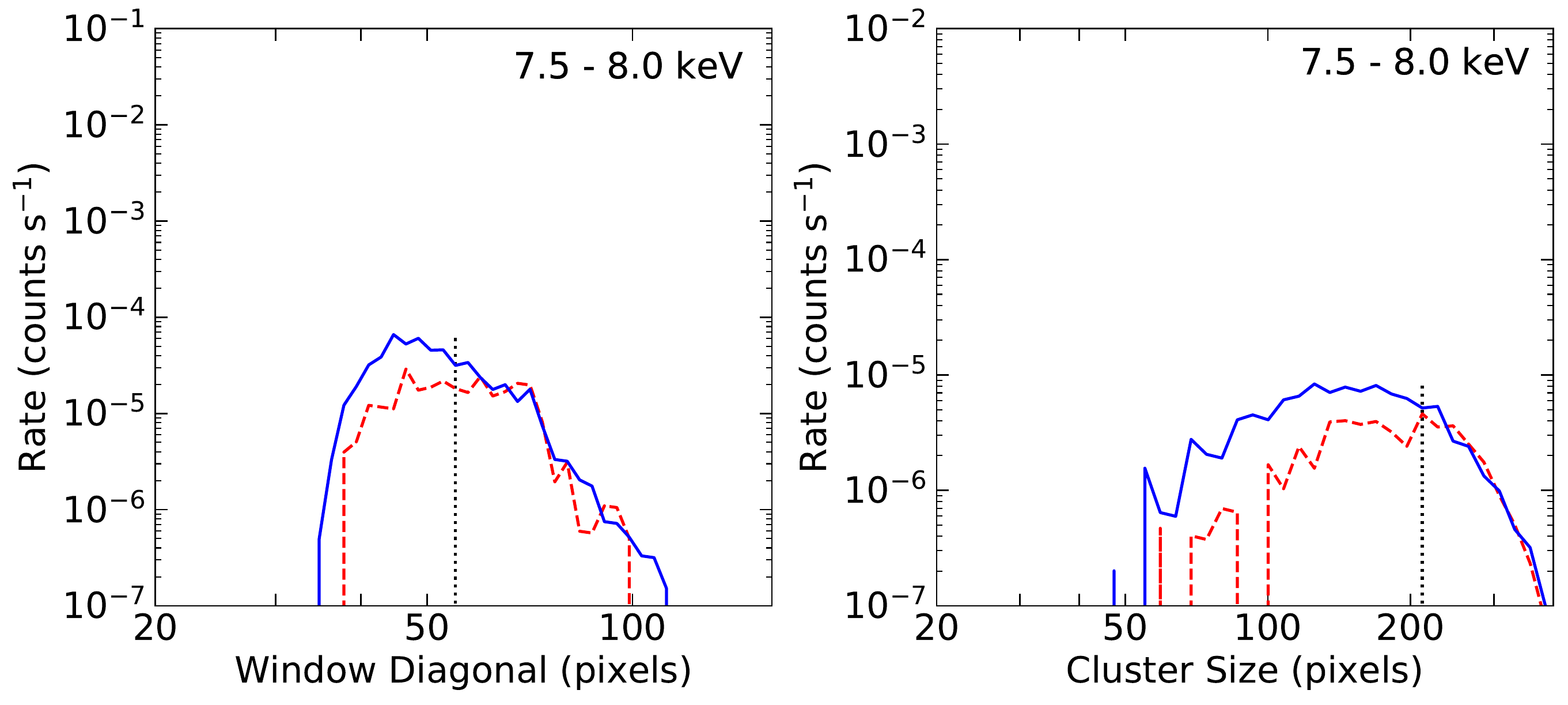}
\caption{One-dimensional distribution of readout window diagonal size ($L$) and cluster size ($S$) in each energy bin from 2 to 8 keV. The solid blue curve indicates the on-source distribution and the dashed red curve indicates the off-source distribution. The dotted black line marks the $L$ or $S$ value (see text for details) above which the on- and off-source distributions are consistent with each other.} 
\label{fig:1d}
\end{figure*}

In order to determine the boundary that separates the source and background distributions,  we extract the 1D distribution along $L$ or $S$ in each energy bin (see Figure~\ref{fig:1d}). The on-source distribution and the off-source distribution are consistent with each other at high $L$ or $S$, suggesting that the background is the dominant component in these parameter spaces. While in low $L$ or $S$, there are significant excesses in the on-source distributions, indicative of source dominance. Therefore, for either $L$ or $S$ in each energy bin, we start from the peak location of the on-source distribution and search toward high $L$ or $S$ values, taking into account Poisson fluctuations, and find the first point where the on-source distribution is consistent with the off-source distribution within 3$\sigma$. We adopt these points as the ``boundaries'' between the source-dominant and background-dominant regimes on the $L-E$ or $S-E$ plane, shown as crosses in Figure~\ref{fig:pure_src}.  We then use empirical functions to fit the boundaries and obtain the discrimination curves.  The function can be in any form. In this case, a quadratic polynomial is used, shown in Figure~\ref{fig:pure_src}. We note that the two examples shown in Figure~\ref{fig:track} are respectively on the two sides of the discrimination curves. 
 
The discrimination curves determined using the above algorithm are a function of the data quality, and may depend on the signal to noise ratio.  We investigate this problem with a portion of the data and find that it is insensitive to the exposure time but the boundaries vary with the source-to-background flux ratio. This is mainly because the source and background distributions overlap on the $L-E$ and $S-E$ planes. Thus, the above curves should be calculated for different sources. 

In addition, laboratory tests suggest that $\sim$99\% of the X-ray events show a single charge island, i.e., all pixels above the noise threshold are in a group of connected pixels. On the other hand, charged particles may leave behind multiple charge islands. This can be taken as another criterion to distinguish source and background events.  We adopt the depth-first search algorithm to identify independent charge islands in the image.

\section{Conclusion and discussion}
\label{sec:diss}

To conclude, background events in the PolarLight data in the energy range of 2--8 keV can be discriminated using the discrimination curves on the $L-E$ and $S-E$ planes (see Figure~\ref{fig:pure_src}), plus a criterion that the number of charge islands is greater than one. 

Applying the above discrimination rules to the off-source observations, we find that $\sim$74\% of data are classified to be background events and $\sim$26\% are source-like events.  These fractions are subject to the choice the discrimination function. As mentioned above, simulations reveal that $\sim$28\% of the background particles result in an energy deposit via a secondary electron in the energy band of our interest~\citep{Huang2021}. These events follow the identical physical processes as a source photon does, and are thus indistinguishable. Thus, the discrimination rules are sufficiently effective and can remove the majority of background events that are removable.  Compared with the algorithm used in \citet{Feng2020a}, we note that the new method no longer utilizes the track eccentricity to screen data, because screening with eccentricity may affect the measurement of polarization and should best be avoided. 

\begin{figure}[t]
\centering
\includegraphics[width=0.4\columnwidth]{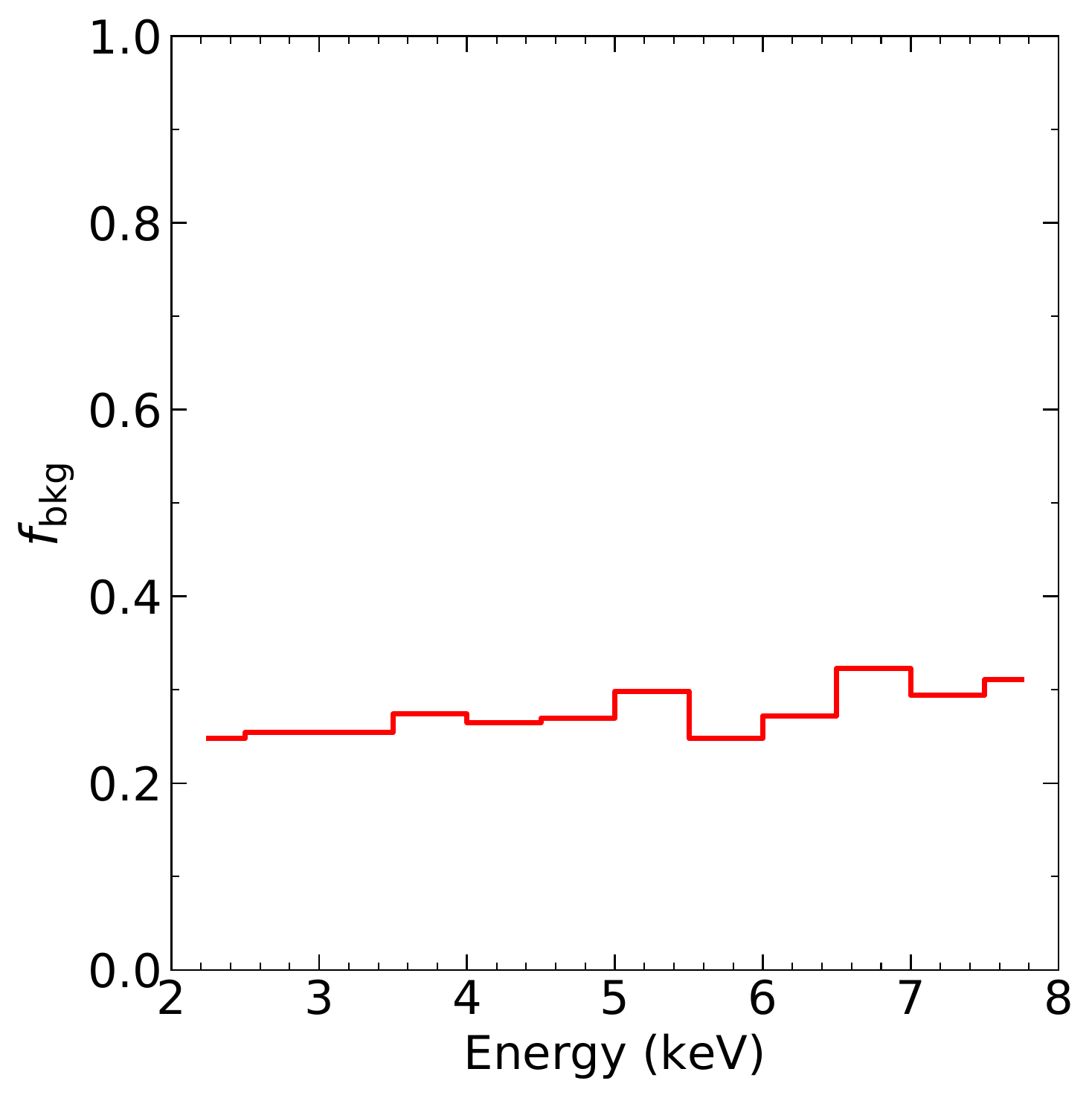}
\includegraphics[width=0.4\columnwidth]{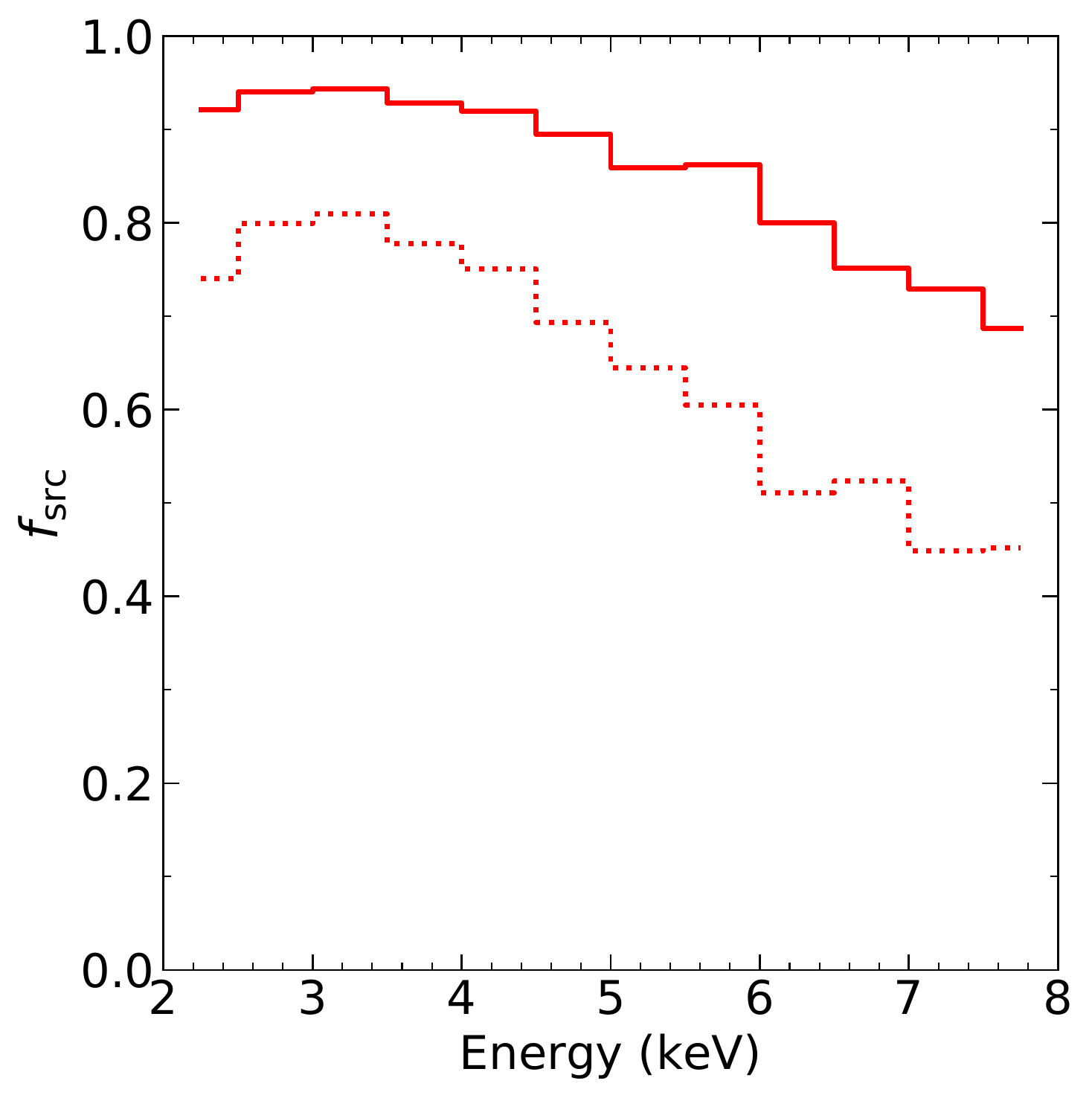}
\caption{\textbf{Left}: fraction of remaining background events after discrimination as a function of energy. \textbf{Right}: fraction of source events as a function of energy before (dotted) and after (solid) background discrimination in the data for the Crab nebula.}
\label{fig:frac}
\end{figure}

The remaining background after discrimination as a function of energy, as well as the source photon fraction for on-source observations of the Crab nebula, are shown in Figure~\ref{fig:frac}. For the on-source observations of the Crab nebula, $\sim$82\% of the events remain after the discrimination. The background fraction is estimated to be $\sim$25\% in the raw data, and reduced to $\sim$8\% after discrimination.  We note that these fractions are accurate to roughly 2\% due to systematic uncertainties in the background determination. This help improve the signal to noise ratio and the instrument sensitivity.  With such a background fraction and an observing time of 1~Ms, PolarLight can reach a minimum detectable polarization of 10\% for a 0.2~Crab source, or 5\% for a 0.5~Crab source. 

The means proposed in this paper does not rely on complicated calculation or analysis. All the quantities (the energy, readout window size, cluster size, and number of charge islands) can be readily extracted from the event images, immune from uncertainties or ambiguities in parameter tuning.  The same method can be applied to observations of other targets.  We note that, as the source and background distributions overlap on the $L-E$ and $S-E$ planes, the discrimination curves depend on the source-to-background flux ratio. The brighter the source is, the higher the discrimination curves are on the $L-E$ and $S-E$ planes. The criterion on the number of charge islands is independent of the source flux.

In the future, the same technology used in PolarLight can be utilized in larger X-ray telescopes, such as the Imaging X-ray Polarimetry Explorer~\citep{Weisskopf2016} and the enhanced X-ray Timing and Polarimetry~\citep{Zhang2019}. These are imaging instruments and suffer less from background contamination thanks to the small focal spot. The means proposed in this paper is still useful to minimize the background level and help improve their sensitivities. 

\begin{acknowledgements}
We thank the referee for useful comments. HF acknowledges funding support from the National Natural Science Foundation of China under the grant Nos.\ 11633003, 12025301 \& 11821303, the CAS Strategic Priority Program on Space Science (grant No.\ XDA15020501-02), and the National Key R\&D Project (grants Nos.\ 2018YFA0404502 \& 2016YFA040080X).
\end{acknowledgements}


\label{lastpage}

\end{document}